\documentclass[prb,aps,onecolumn,showpacs,floats,a4paper,times,12pt]{revtex4}
\usepackage{epsfig,bm,epsf,graphics}
\begin{document}
\draft
\title{Spin Resistivity in Frustrated Antiferromagnets}
\author{Y. Magnin$^a$, K. Akabli$^{a,b}$, and H. T. Diep$^a$\footnote{ Corresponding author, E-mail: diep@u-cergy.fr }}
\address{
$^a$ Laboratoire de Physique Th\'eorique et Mod\'elisation,
Universit\'e de Cergy-Pontoise, CNRS, UMR 8089\\
2, Avenue Adolphe Chauvin, 95302 Cergy-Pontoise Cedex, France\\
$^b$ Department of Physics, Okayama University,  3-1-1 Tsushima-naka, Okayama 700-8530, Japan.\\}

\begin{abstract}

In this paper we study the spin transport in frustrated antiferromagnetic FCC films by Monte Carlo simulation. In the case of Ising spin model, we show that the spin resistivity versus temperature exhibits a discontinuity at the phase transition temperature:  an upward jump or a downward fall, depending on how many parallel and antiparallel localized spins interacting with a given itinerant spin. The surface effects as well as the difference of two degenerate states on the resistivity are analyzed.   Comparison with  non frustrated antiferromagnets is shown to highlight the frustration effect.  We also show and discuss the results of the Heisenberg spin model on the same lattice.

\end{abstract}

\maketitle

\section{Introduction}

The resistivity in magnetically ordered materials has been studied experimentally and theoretically during the last 50 years.  Unlike the resistivity in non magnetic systems which is due mainly to the scattering of conduction electrons by phonons, the resistivity in magnetic materials depends on the magnetic ordering.  It has been theoretically shown that in ferromagnets the resistivity is due to the spin-spin correlation by several workers\cite{DeGennes,Fisher,Haas,Kataoka}, using various approximations such as mean-field theories and Boltzmann's equation.  The main difference of these treatments resides on the way the correlation length is taken into account: long-range correlation gives rise to a divergence of the magnetic resistivity $\rho$ at the transition temperature $T_C$, while short-range correlation causes a rounded peak.  Experimental data show that $\rho$ has several forms depending on the materials: $\rho$ shows a peak, a rounded shoulder, or just a change of slope at $T_C$. In the last case, it is $d\rho /dT$ which shows a peak\cite{Craig,Shwerer,Matsukura}.   In recent experiments performed on different kinds of magnetic pure or doped materials ranging from insulators, semiconductors to superconductors,   the form of $\rho$ is very different\cite{Xia,Wang-Chen,Santos,Li,Lu,Du,Zhang,McGuire}. These theories and experiments suggest that the shape of $\rho$ depends on the different magnetic interactions in the system, and on the local magnetic ordering through which the itinerant spins evolve.  Recently, the study of spin resistivity has attracted again much attention  due to numerous applications since the discovery of the so-called giant magnetoresistance\cite{Baibich,Grunberg,Fert,review,Dietl,Barthe,Wysocki}.

Let us recall first some results in the ferromagnetic case.  At very low temperatures $T$ where spin-wave scattering is dominant, the resistivity has been studied in detail showing a $T^2$ behavior\cite{Kasuya,Turov}. Near the transition, we have shown in  recent papers\cite{Akabli2,Akabli} that the pronounced peak  at the Curie temperature $T_C$ of the magnetic resistivity of ferromagnets can be explained by the scattering of itinerant spins by defect clusters formed in the lattice around $T_C$. This picture has an advantage: it can be checked easily by calculating the numbers and the sizes of the clusters numerically during the simulation using the Hoshen-Kopelmann's algorithm\cite{Hoshen}.  Note that  clusters of down spins can be considered as magnetic impurities embedded in a up-spin sea which have been theoretically studied by Zarand et al.\cite{Zarand} where they found also a pronounced peak of the resistivity.  Of course, the cluster sizes reflect the correlation length used in early theories\cite{DeGennes,Haas,Kataoka}. It is not a surprise that our results for ferromagnets were in agreement with all these theories.

The case of antiferromagnets has not been well studied. There were only a few works which mentioned briefly some behaviors. Let us cite the work by Haas where he stated that in antiferromagnets there is no peak in $\rho$ using the spin-spin correlation in the Boltzmann's equation\cite{Haas}. Our recent works\cite{Magnin,Akabli3} on  the simple cubic and the body-centered cubic Ising antiferromagnets show that there is indeed no peak in $\rho$:  $\rho$ varies with $T$ in a manner similar to that of the internal energy versus $T$. As a consequence, the differential resistivity $d\rho/dT$ shows a peak at the transition temperature $T_C$ just like the specific heat.  Interestingly enough, this behavior has been experimentally observed in MnSi and related compounds\cite{Stishov,Stishov2}.  We will  show below some results of these non frustrated antiferromagnets for comparison with the frustrated case studied in this paper.

Our purpose  is to show in this paper one of the remarkable cases: the face-centered cubic (FCC) antiferromagnet (AF).  This system is known to be fully frustrated with a strong first-order transition in the Ising case\cite{Diep2005,Pham2009}.   The spin resistivity is known to be very sensitive to the nature of the ordering of the media through which the itinerant spins move: local disordering (for instance, disordering near film surfaces, around magnetic impurities), magnetic instability, ...  The FCC AF is thus a very good candidate where exotic behaviors are expected for the spin resistivity.  This will indeed be seen in this work.
By using Monte Carlo (MC) simulation, we show that frustration and interaction range between itinerant spins and lattice spins play a crucial role on the spectacular discontinuity of the magnetic resistivity at $T_C$.   In the case of Heisenberg spin, the transition is also of first order, though weaker, in the bulk\cite{DiepFCC}.   We  study also this model in this work to outline the effect of spin continuous degrees of freedom on the resistivity.

The paper is organized as follows. Section II is devoted to the description of the model and the calculation method. In section III, we show MC results on the temperature dependence of the magnetic resistivity in the Ising spin model.  Section IV is devoted to results of the Heisenberg case. Discussion and explanation are given with regard to the  transport mechanism. Concluding remarks are given in section IV.

\section{Model and Simulation Procedure}
\subsection{Model}
We consider a thin film of FCC lattice structure where each lattice site is occupied by an Ising spin whose values are $\pm 1$. The Heisenberg spin model is considered in section \ref{Heis}.  Interaction between the lattice spins is limited to nearest-neighbor (NN) pairs with
the following Hamiltonian :
\begin{eqnarray}
\mathcal{H}_l & = & -\sum_{(i,j)}J_{i,j}\vec{S}_{i}.\vec{S}_{j}\label{HamilR}
\end{eqnarray}
where $\vec S_i$ is an Ising spin, $J_{i,j}$ the exchange integral between the NN spin pair $\vec S_i$ and $\vec S_j$.  Hereafter we take $J_{i,j}=J_s$  for surface spins and $J_{i,j}=J$ for other NN spin pairs. We consider here the antiferromagnetic interaction $J_s,J<0$ for the rest of this paper.   The system size is $N_x \times N_y \times N_z$ where $N_x$ is the number of FCC cells in the $x$ direction etc. Periodic boundary conditions (PBC) are used in the $x$ and $y$ directions while the surfaces perpendicular to the $z$ axis are free. The film thickness is $N_z$.

\begin{figure}[!h]
\begin{center}
\includegraphics[width=5cm]{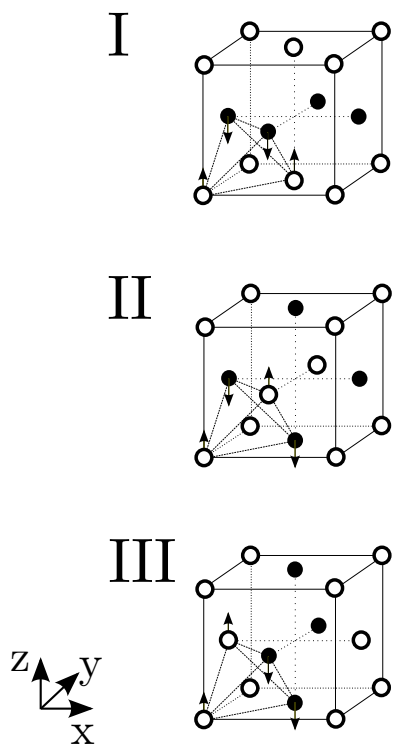}
\caption{Ground state spin configuration of the FCC cell at the film surface (basal $xy$ plane).
The horizontal (vertical) axis is the $x$ ($z$) axis.
Upper: ground state when $|J_s|<0.5|J|$, Middle and Lower: first and second ground states when $|J_s|>0.5|J|$. }
\label{deg_fcc}
\end{center}
\end{figure}

The  FCC AF is a fully frustrated system which is composed of tetrahedra each of which has
four equilateral triangles.  We know that it is impossible to fully satisfy
simultaneously the three antiferromagnetic bond interactions on each triangle. As a consequence,
the bulk lattice has an infinite ground-state degeneracy\cite{Diep2005}. In the case of a thin film,
the surface spin configuration depends on $J_s$ as shown in  Fig. \ref{deg_fcc}\cite{Pham2009}.


For $|J_s|<0.5|J|$, the ground state is composed of ferromagnetic $xy$ planes antiferromagnetically stacked in the $z$ direction a shown in the upper figure of Fig. \ref{deg_fcc}.
For $|J_s|>0.5|J|$, the ground state is two-fold degeneracy as shown in the middle and lower figures of Fig. \ref{deg_fcc}.   The difference of these two configurations is that the middle figure is an alternate stacking of up- and down-spin planes in the $y$ direction while the lower figure is an alternate stacking of up- and down-spin planes in the $x$ direction.   These degenerate states are not equivalent in the spin transport in the $x$ direction as seen below: in the first degenerate state, the itinerant spins move in the $x$ direction between an up-spin plane and a down-spin plane, while in the second degenerate state the itinerant spins meet successively an up-spin plane and a down-spin plane perpendicular to their trajectories.  We will present our results for these two cases separately.

\subsection{Multi-step Averaging}
The procedure of our simulation can be split into two steps. The first step consists in equilibrating the lattice at a given temperature $T$ without itinerant electrons.  When equilibrium is reached, we study thermodynamic properties of the film so as to determine its Neel temperature by examining  quantities like internal energy, specific heat, susceptibility and magnetization as functions of $T$\cite{Metro,Binder}.

In the second step, we randomly add $N_0$ polarized itinerant spins into the thermalized FCC lattice. In the structure, each itinerant electron interacts with lattice spins in a sphere of radius $D_1$ centered on its position, and with other itinerant electrons in a sphere of radius $D_2$. We define these interactions as follows
\begin{eqnarray}
\mathcal{H}_r & = & -\sum_{i,j}I_{i,j}\vec{\sigma}_i .\vec{S}_j
\end{eqnarray}
where $\sigma_i$ is the Ising spin of itinerant electron and $I_{i,j}$ denotes the interaction that depends on the distance between an electron $i$ and the spin $\vec{S}_j$ at the lattice site $j$. We use the following interaction expression :
\begin{eqnarray}
I_{i,j} & = & I_{0}e^{-\alpha r_{ij}} \mbox{\hspace{0,3cm}with\hspace{0,3cm}} r_{ij}=|\vec{r}_i-\vec{r}_j|
\end{eqnarray}
where $I_0$ and $\alpha$ are  constants which will be chosen in section \ref{choice}. In the same way, interaction between itinerant electrons is defined by :
\begin{eqnarray}
\mathcal{H}_m & = & -\sum_{i,j}K_{i,j}\vec{\sigma}_i .\vec{\sigma}_j\\
K_{i,j} & = & K_{0}e^{-\beta r_{ij}}\label{K}
\end{eqnarray}
with $\sigma_i$ the spin of itinerant electron and $K_{i,j}$ the interaction that depends on the distance between electrons $i$ and $j$.  The choice of the constants $K_0$ and $\beta$ is discussed in \ref{choice}.\\
Dynamics of itinerant electrons is ensured by an electric field applied along the $x$ axis. Electrons enter the system at the first end, travel in the $x$ direction, leave the system at the second end. The PBC on the $xy$ planes ensure that the electrons who leave the system at the second end are to be reinserted at the first end. For the $z$ direction, we use the mirror reflection at the two surfaces.  These boundary conditions  are used in order to conserve the average density of itinerant electrons.  One has
\begin{eqnarray}
\mathcal{H}_E & = & -e\vec{\epsilon}.\vec{\ell}
\end{eqnarray}
where $e$ is the charge of electron, $\vec \epsilon $ the applied electrical field and $\vec \ell$ the displacement vector of an electron.\\
Since the interaction between itinerant electron spins is attractive,  we need to add a chemical potential in order to avoid a possible collapse of electrons into some points in the crystal and to ensure a homogeneous distribution of electrons during the simulation. The chemical potential term is given by
\begin{eqnarray}
\mathcal{H}_c & = & D\vec{\nabla}_rn(\vec{r})\label{pot}
\end{eqnarray}
where $n(\vec r)$ is the concentration of itinerant spins in the sphere of $D_2$ radius, centered at $\vec r$. $D$ is a constant parameter appropriately chosen.\\

The procedure of spin dynamics can be described as follows. After injecting $N_0$ itinerant electrons in the equilibrated antiferromagnetic FCC lattice, we calculate the energy $E_{old}$ of an itinerant electron taking into account all interactions described above. Then we perform a trial move of length $\ell$ taken in an arbitrary direction with random modulus in the interval $[R_1,R_2]$ where $R_1=0$ and $R_2=a/\sqrt{2}$ (nearest-neighbor distance),  $a$ being the lattice constant.  Note that the move is rejected if the electron falls in a sphere of radius $r_0$ centered at a lattice spin or at another itinerant electron. That excluded space emulates the  Pauli exclusion. We calculate the new energy $E_{new}$ and use the Metropolis algorithm to accept or reject the electron displacement.
We choose another itinerant electron and begin again this procedure.  When all itinerant electrons are considered, we say that we have made a MC sweeping, or one MC step. We have to repeat a large number of MC steps to reach a stationary transport regime. When the stationary regime is reached, we perform the averaging to determine physical properties such as magnetic resistivity, electron velocity, energy etc. as functions of temperature.

We emphasize here that in order to have sufficient statistical averages on microscopic states of  both the lattice spins and the itinerant spins, we use the following procedure: after averaging  the resistivity over $N_1$ MC steps we thermalize again the lattice with $N_2$ steps, then equilibrate the itinerant spins with $N_3$ steps before taking back the averaging of the resistivity during $N_1$ steps. We repeat this cycle $(N_1+N_2+N_3)$ for $N_4$ times (typically $N_4$=100).  The total MC steps is therefore equal to $N_4\times (N_1+N_2+N_3)$.   As will be seen below, this procedure reduces  strongly thermal fluctuations observed in our previous work\cite{Akabli2}.  The transport averaging is made for $N_2=100$ configurations of lattice spins. At each configuration, we adjust $N_1$ so that the spin resistivity is calculated during $1000$ lattice sweepings per itinerant spin (each electron passes through the system 1000 times).  $N_2$ depends on $T$: it can be several thousands near $T_C$.  In all, at each $T$ the initial equilibration time for lattice spins lies around $10^5$-$10^6$ steps per spin and statistical averages are made with about $4\times 10^5$ steps per spin ($N_4 \times N_1$).

We define resistivity $\rho$ as :
\begin{eqnarray}
\rho & = & \frac{1}{n_{e}}
\end{eqnarray}
where $n_{e}$ is the number of itinerant electrons crossing a  unit slice perpendicular to the $x$ direction per unit of time.

In this paper we use the lattice size $N_x=N_y=20$ and $N_z=8$.

For studying the spin transport, we consider $N_0=(N_x\times N_y\times N_z)/2$ itinerant spins (one electron per  two FCC unit cells).  Except otherwise stated, we choose interactions $I_0=K_0=0.5$, $D_1 \in [0.6a;2a]$, $D_2=a$, $D=0.35$, $\epsilon=1$, $N_0=1600$, and $r_0=0.05a$.  A discussion on the effect of a variation of each of these parameters will be given in \ref{choice}.

\begin{figure}[!h]
\begin{center}
\includegraphics[width=12cm]{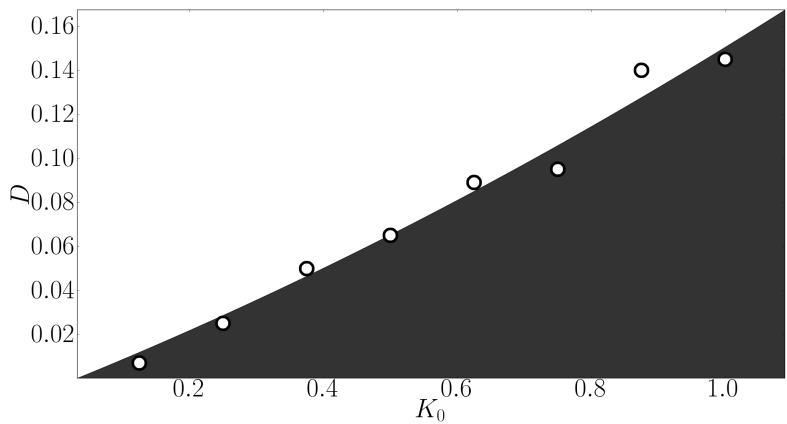}
\caption{Collapse phase diagram in the space $(K_0,D)$. The black zone is the collapse region. $D_1=D_2=a$. See text for comments.}
\label{col}
\end{center}
\end{figure}

Note however that, due to the form of the interaction given by Eq. (\ref{K}),  the itinerant spins have a tendency to form compact clusters to gain energy.  This tendency is neutralized more or less by the concentration gradient term, or chemical potential, given by  Eq. (\ref{pot}).  The value of $D$ has to be chosen so as to avoid a collapse of itinerant spins.  We show in Fig. \ref{col} the phase diagram in the space ($K_0,D$).  The limit depends of course on the values of $D_1$ and $D_2$.

\subsection{Choice of different parameters}\label{choice}
We will show below results obtained for typical values of parameters.  The choice of the parameters has been made after numerous test runs.  Let us describe the principal requirements for the choice:

i) we choose the interaction between lattice spins as unity, i. e. $|J|=1$.

ii) we choose interaction between an itinerant and its surrounding lattice spins so as its energy $E_i$ in the low $T$ region is the same order of magnitude with that between lattice spins. To simplify, we take $\alpha=1$.

iii) interaction between itinerant spins is chosen so that this contribution to the itinerant spin energy is smaller than
$E_i$ in order to highlight to effect of lattice ordering on the spin current. To simplify, we take $\beta=1$.

iv) the choice of $D$ is made in such a way to avoid the formation of  clusters of itinerant spins (collapse) due to their attractive interaction [Eq. (\ref{K})] as shown above.

v) the electric field is chosen not so strong in order to avoid its dominant effect that would mask the effects of thermal fluctuations and of the magnetic ordering.

vi) the density of the itinerant spins is chosen in a way that the contribution of interactions between themselves is neither so weak nor so strong with respect to $E_i$.

Within these requirements, a variation of each parameter does not change qualitatively the results shown below. As will be seen, only the variation of $D_1$ does change drastically the results. That is the reason why we will study in detail the effect of this parameter.  For larger densities of itinerant spins, the resistivity is larger as expected because of additional scattering process between itinerant spins.


In view of the above requirements, we take for the simulations:  $J=-1$ (AF interaction), $I_0=K_0=0.5|J|$, $D_1 \in [0.6a;2a]$, $D_2=a$, $D=0.35$, $\epsilon=1$, $N_0=1600$, and $r_0=0.05a$, $a$ being the FCC lattice constant.   Within these choices, the results in the following will be presented in  the following units: the spin energy is in the unit of $|J|$, the temperature is in the unit of $|J|/k_B$, the distance is in the unit of $a$.

Finally, we keep $\alpha$ constant when varying $D_1$. This is because varying $D_1$ means we include or not include some far neighbors.  If the interaction of these far neighbors follows the same Eq. (3) as the one between shorter neighbors then it is known theoretically that no new interesting physical can occur except the modification of non universal values such as the critical temperature.  This case corresponds to ferromagnetic interaction where further neighbors do not cause interesting effect. On the other hand, when further neighbor interaction is in competition with the interaction of nearer neighbors, the system can be frustrated, then physical properties can radically vary. This is the case studied here.   Now varying $\alpha$ to keep the the interaction constant will not change observed physical behaviors. What will change is the relative value of the energy and therefore the value of the transition temperature but not the qualitative behavior of the system.

\section{Results for the Ising case}
\begin{figure}[!h]
\begin{center}
\includegraphics[width=9cm]{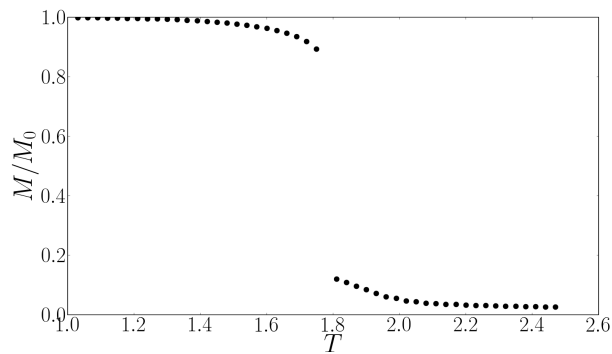}
\caption{Staggered magnetization of antiferromagnetic FCC thin
film of thickness $N_z=8$ versus $T$. The transition temperature $T_C\simeq 1.79$.}
\label{m_plot}
\end{center}
\end{figure}

We show in Fig. \ref{m_plot} the staggered magnetization of the lattice as a function of $T$.
As seen here the transition is of first order with a discontinuity at $T_C\simeq 1.79$.  Note that the Ising AF FCC thin film shows a first-order transition down to a thickness of about four atomic layers\cite{Pham2009}.


\subsection{Resistivity in the first degenerate state}\label{IDG1}

 We consider the first degenerate configuration shown in the middle figure of Fig. \ref{deg_fcc} with
 $J_s=J=-1$.
 To understand the behavior of the spin  resistivity which will be shown below, let us first show how various physical quantities at a given temperature depends on $D_1$.

 In the ferromagnetic state,  increasing (decreasing)  $D_1$ results in an increase (decrease) of the number of "parallel" lattice spins which interact with the itinerant spin. This means that increasing (decreasing) $D_1$ results in a decrease (increase) of the energy of the itinerant spin. In antiferromagnets, the situation is different: changing $D_1$ will result in an oscillatory change of the difference of the numbers of parallel and antiparallel spins in the sphere of radius $D_1$, namely $\Delta N_{\uparrow \downarrow}=N_{\uparrow}-N_{\downarrow}$.  This is because of antiferromagnetic ordering.

\begin{figure}[!h]
\begin{center}
\includegraphics[width=12cm]{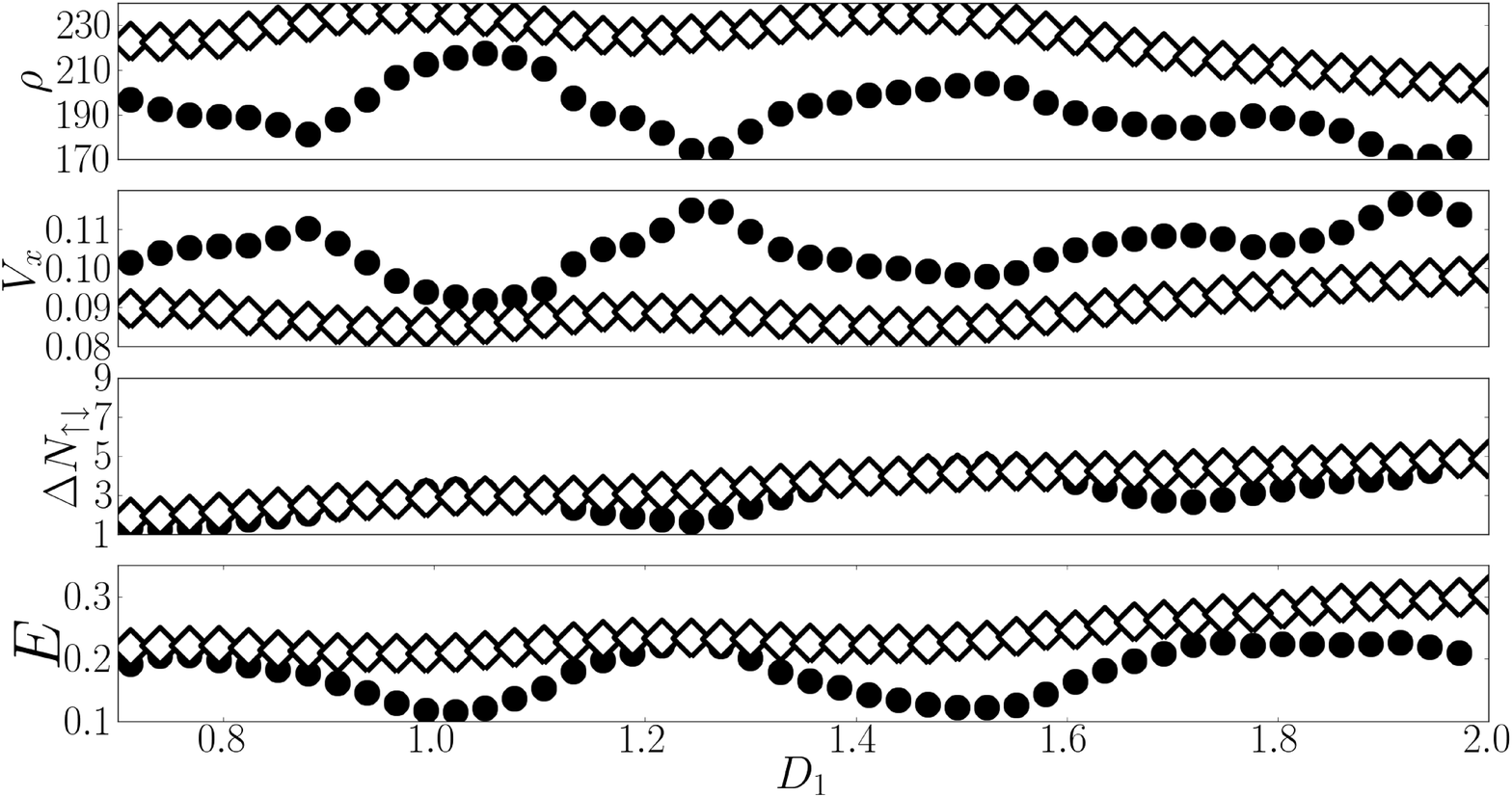}
\caption{Different physical quantities versus $D_1$ in unit of the lattice constant $a$ in the case of the first degenerate configuration.  From top to bottom:  Resistivity,
velocity on the $x$ axis, difference of up- and down-spin numbers,
energy of an itinerant spin.  In each plot circles
corresponds to  $T=1.65$, and diamonds to $T=2$.
 $N_z=8$, $N_0=1600$,
$J_s=J=-1.0$ and $D=0.35$.}
\label{ev_plot}
\end{center}
\end{figure}

We show in Fig. \ref{ev_plot} the resistivity, the spin velocity in the $x$ direction,
$\Delta N_{\uparrow \downarrow}$ and the energy of an itinerant spin at $T=1.65$  below the transition  and at $T=2$ in the paramagnetic lattice, for different values of $D_1$.  The following remarks are in order:
\begin{itemize}
\item  As said above, at low $T$, $\Delta N_{\uparrow \downarrow}$ oscillates with varying $D_1$.
When $\Delta N_{\uparrow \downarrow}$ is maximum, i. e. the number of up spins is large, the energy of the itinerant spin is low. As a consequence the itinerant spin will not move easily under the electric field, its velocity is therefore slowed down, making the resistivity to increase.

\item At high $T$, the lattice spins are disordered, there are no more shells alternately of up spins and down spins around an itinerant spin.  So the oscillatory behavior of $\Delta N_{\uparrow \downarrow}$ is reduced as seen in   Fig. \ref{ev_plot} at $T=2$.
\end{itemize}


\begin{figure}[!h]
\begin{center}
\includegraphics[width=12cm]{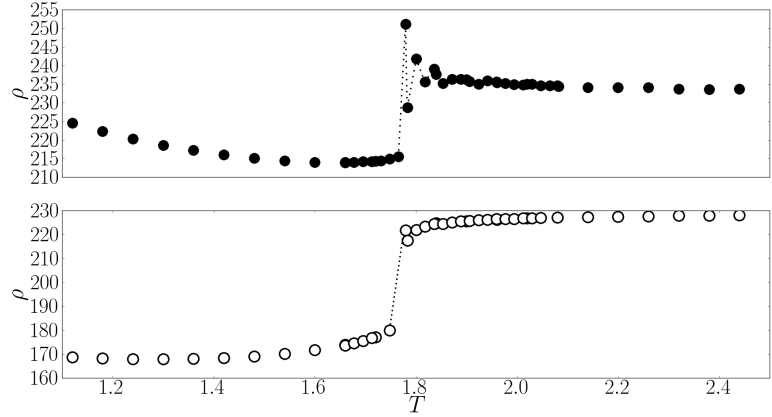}
\caption{Resistivity of thin film of size  $N_x=N_y=20$ and $N_z= 8$ for $N_0=1600$
itinerant spins versus $T$ for $D_1=a$ (black circles) and $D_1=1.25a$ (white circles), $a$ being the lattice constant. Case  of the first degenerate
state. $J_s=J=-1.0$, $I_0=K_0=0.5$, $D=0.35$.}
\label{r_plot}
\end{center}
\end{figure}

 In Fig. \ref{r_plot} we show the spin resistivity $\rho$ versus $T$ for two typical values of $D_1$. In all cases resistivity $\rho$ is small for low $T$ then increases with increasing $T$.  At $T_c$, it undergoes a discontinuity upward jump. After transition, the resistivity decreases slowly to the same value for all $D_1$ in paramagnetic phase.  We explain the behavior of $\rho$ at different temperature regions:

\begin{itemize}
\item When $T \rightarrow 0$, the resistivity slightly increases because itinerant spins search to minimize energy by occupying low-energy positions in the periodic lattice. Since thermal energy and electric field are not strong enough to make them move, the itinerant spins are somewhat frozen in some almost periodic positions, namely a pseudo crystallization.  We have studied the spatial distribution of itinerant spins. The results show indeed a radial distribution with peaks up  to rather long-range positions at low $T$, namely up to 4th nearest neighbors.  Note that the increase of $\rho$ when $T \rightarrow 0$ has been observed in many experiments among which we can mention: Fig. 11 of the paper by Chandra et al. on CdMnTe,\cite{Chandra}
 Fig. 2 of the paper by Du et al. for MnFeGe,\cite{Du} Fig. 6a of the paper by McGuire et al. on AF superconductors LaFeAsO,\cite{McGuire} Fig. 2 of the paper by Lu et al. on AF LaCaMnO,\cite{Lu} and Fig. 7 of the paper by Santos et al. on AF LaSrMnO.\cite{Santos}
 Note however that most of these experiments concern doped semiconductors.  In semiconductors, the carrier concentration increases with increasing $T$. Our model has a number of itinerant spins which is independent of $T$ in each simulation.    So we cannot compare quantitatively our results with experiments on semiconductors, in particular these latter are often magnetically disordered systems.  But in each simulation, we can take another concentration (see our previous paper\cite{Akabli2}): the results show that the resistivity is somewhat modified but keeps the same feature, except the fact that the stronger the concentration is the smaller the peak at $T_C$ becomes. Therefore, we believe that some generic effects independent of carrier concentration will remain.

Note also that our results of the resistivity at low $T$ depends on our model. This behavior ($\rho$ increases with decreasing $T$) is observed only when we introduce a rather strong interaction between itinerant spins (variable $K_0$).  Reducing $K_0$ will suppress this tendency.

On the hypothesis of frozen electrons, there is a reference on the charge-ordering at low $T$ in Pr$_{0.5}$Ca$_{0.5}$MnO$_3$\cite{Zhang} due to some strain interaction. A magnetic field can make this ordering melted giving rise to a depressed resistivity. Though our model does not correspond to this material, the fundamental concept is similar.  For the system
Pr$_{0.5}$Ca$_{0.5}$MnO$_3$, which shows commensurate charge order, the "melting"
fields at low temperatures are high, on the order of 25 Tesla\cite{Zhang}.  

We mention here that low-$T$ behaviors can be also studied by alternative Kubo and Landauer methods as shown by Ref. 22.

    When $T$ increases, thermal energy unfreezes itinerant electrons, the system is progressively unfrozen and the resistivity slightly decreases and then increases up to  $T_C$.
\item At $T_C$, $\rho$ exhibits a discontinuity due to the discontinuity of the lattice magnetization to which the itinerant spins are coupled.  For the first degenerate configuration (Fig. \ref{deg_fcc}, middle) $\rho$ makes an upward jump.
\item After $T_C$, the lattice is paramagnetic:  there is no significant effect of $D_1$ as discussed earlier.
\end{itemize}


\begin{figure}[!h]
\begin{center}
\includegraphics[width=12cm]{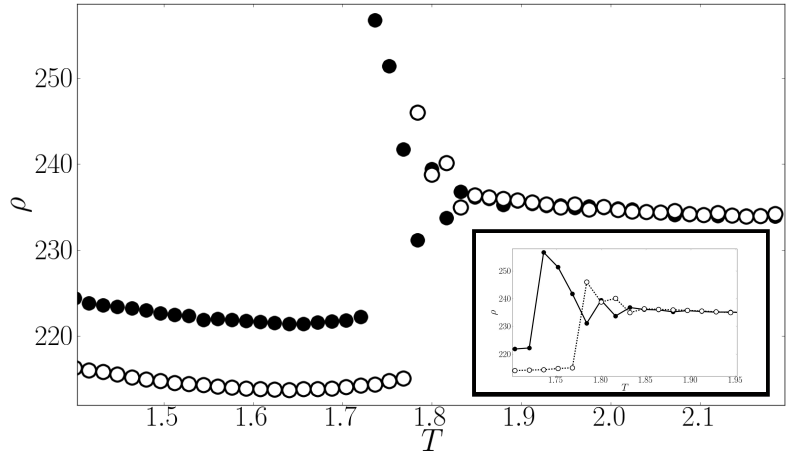}
\caption{Resistivity versus $T$ for two values of magnetic field $B$ with
 $D_1=a$, $N_z=8$, $N_0=1600$ and  $J_s=J=-1.0$. Black circles correspond to $B=0.75$
 and white circles to $B=0.25$.}
\label{RB}
\end{center}
\end{figure}

We show now the effect of the magnetic field in Fig. \ref{RB}.  We observe here that the peak hight increases with increasing $B$, contrary to the case of ferromagnets where the peak diminishes with increasing $B$\cite{Akabli2,Magnin}. The difference can be explained by the fact that in antiferromagnets the magnetic field causes a transition at some $T$ by returning antiparallel spins and thus enhances critical fluctuations while in ferromagnets the magnetic field suppresses fluctuations and forbids a phase transition.  Since the peak height is proportional to critical fluctuations, it is not surprising that the peak increases with increasing $B$ in antiferromagnets.  Note that $T_C$ diminishes with increasing $B$ as expected in antiferromagnets.


\subsection{Resistivity in the second degenerate state}\label{IDG2}

\begin{figure}[!h]
\begin{center}
\includegraphics[width=12cm]{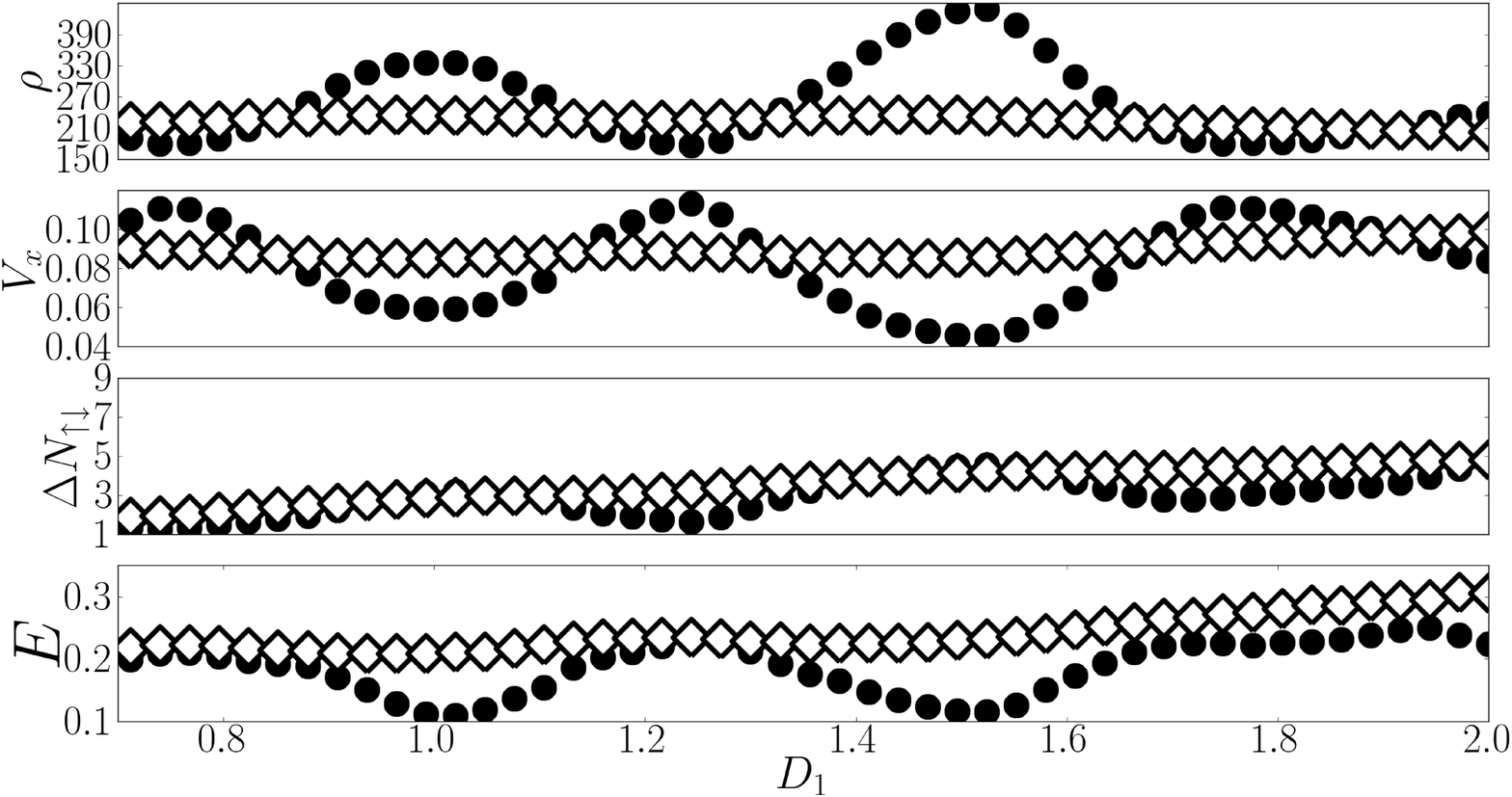}
\caption{Resistivity, spin velocity, $\Delta N_{\uparrow \downarrow}$ and energy of itinerant spin versus $D_1$ (in unit of the lattice constant $a$) at temperatures
 $T=1.65$ (black circles) and $T=2.0$ (white diamonds) in the case of second degenerate
configuration. $N_z=8$, $N_0=1600$, $J_s=J=-1.0$.}
\label{RD1-DG2}
\end{center}
\end{figure}

\begin{figure}[!h]
\begin{center}
\includegraphics[width=12cm]{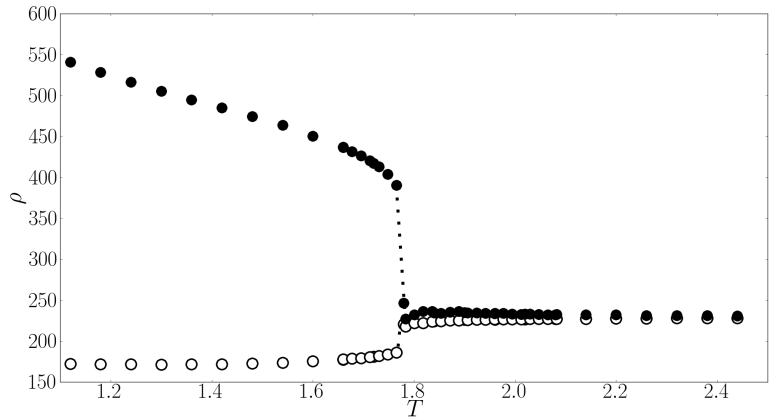}
\caption{Resistivity versus temperature in the case of second
degenerate state for $D_1=a$ (black circles) and $D_1=1.25a$
(white circles) with $N_z=8$, $N_0=1600$, $J_s=J=-1.0$, $I_0=K_0=0.5$, $D=0.35$.}
\label{RT-DG2}
\end{center}
\end{figure}

Let us consider the second degenerate configuration where the ferromagnetic up- and down-spin planes are perpendicular to the spin current in the $x$ direction (lower figure in Fig. \ref{deg_fcc}).  We show first in Fig. \ref{RD1-DG2} the resistivity, the electron velocity, $\Delta N_{\uparrow \downarrow}=N_{\uparrow}-N_{\downarrow}$ and the energy of an itinerant spin at two temperatures,
below and above $T_C$, as functions of $D_1$.  One observes here a cross-over of low-$T$ and high-$T$ resistivities at different positions of $D_1$.  So, depending on $D_1$, low-$T$ resistivity  can be smaller or larger than that of high-$T$.  At $T_C$, $\rho$ can jump upward or downward depending on the value of $D_1$.  This is what we see in  Fig. \ref{RT-DG2}.    Note that, as in the case of the first degenerate configuration, a minimum energy of itinerant spin corresponds to a minimum of the velocity and a maximum of the resistivity.  These quantities are closely related to each other as expected from the physical picture described above.

We can approximately identify the intervals of $D_1$ where $\rho$ jumps (falls) by looking at the top panel of Fig. \ref{RD1-DG2}: at a given $D_1$, $\rho$ jumps (falls) when $\rho$ at $T=1.65$ is lower (higher) than that at $T=2$.

\begin{figure}[!h]
\begin{center}
\includegraphics[width=12cm]{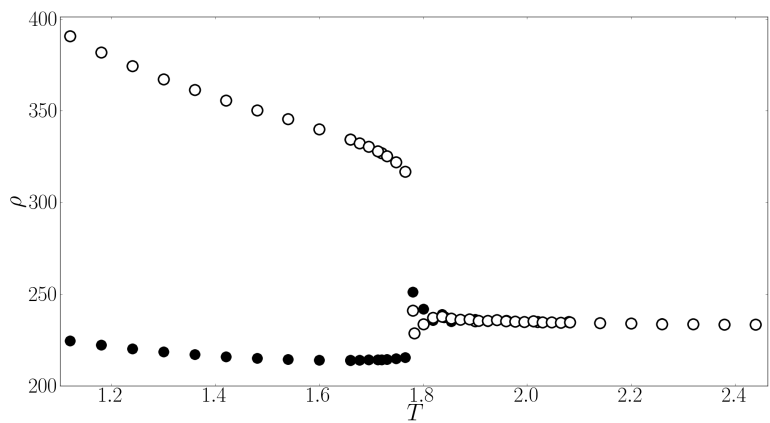}
\caption{Resistivity versus $T$ for first (black points) and second degenerate (white points) configurations for $D_1=a$ with $N_z=8$, $N_0=1600$, $J_s=J=-1.0$,  $I_0=K_0=0.5$, $D=0.35$.}
\label{DG1-DG2}
\end{center}
\end{figure}

\subsection{Surface Effects}
\begin{figure}[!h]
\begin{center}
\includegraphics[width=12cm]{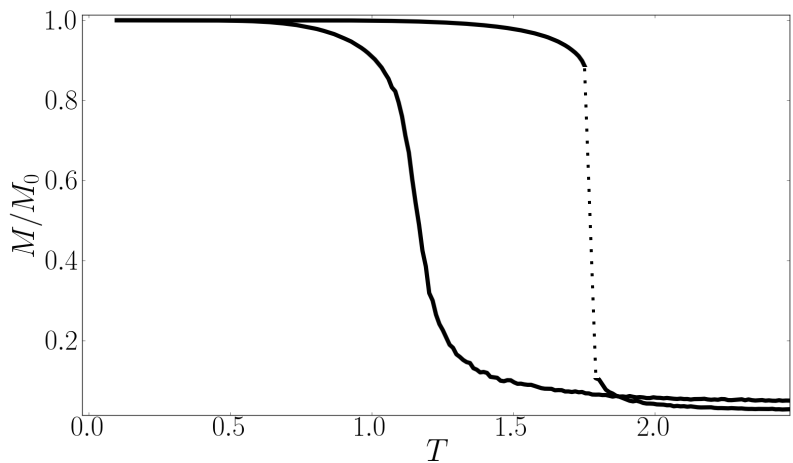}
\caption{Layer magnetizations versus $T$ for $J_s=J_p=-0.5$ and $J=-1$. Other parameters: $D_1=a$, $N_z=8$,
$N_0=1600$,  $I_0=K_0=0.5$, $D=0.35$.  The surface transition is at $T_1\simeq 1.2$. The vertical dotted line is a guide to the eye indicating the discontinuous fall of interior layer magnetization. }
\label{SM}
\end{center}
\end{figure}

\begin{figure}[!h]
\begin{center}
\includegraphics[width=12cm]{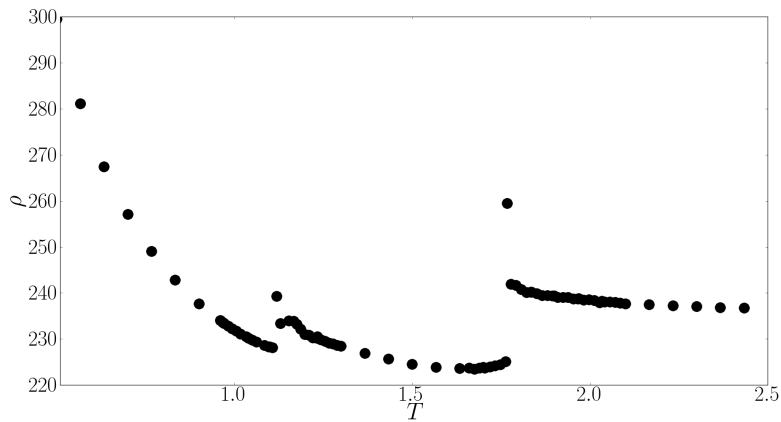}
\caption{Resistivity versus temperature $T$ in the case shown in Fig. \ref{SM}.  There are two anomalies occurring respectively at the surface transition temperature and at the bulk one.  }
\label{RT-SE}
\end{center}
\end{figure}

In order to enhance the surface effect, in addition to a small value of $J_s$ we allow the exchange interaction between a surface spin and its neighbors in the beneath layer to be $J_p$ which will be taken to be small in magnitude.  We show in Fig. \ref{SM} the surface magnetization and the magnetizations of the interior layers as functions of $T$ for $J_s=J_p=-0.5$ and $J=-1$.  As seen here, the surface transition takes place at a lower temperature $T_1\simeq 1.2$ while interior layers become disordered at $T_2\simeq 1.8$.  As a consequence, one expects that the surface fluctuations at $T_1$ will induce an anomaly in $\rho$ in addition to that at $T_2$. This is shown in Fig. \ref{RT-SE}.  Note that the increase of $\rho$ at low $T$ is an effect of a pseudo crystallization of itinerant spins at low $T$ as discussed above.


\subsection{Traveling paths}
\begin{figure}[!h]
\begin{center}
\includegraphics[width=12cm]{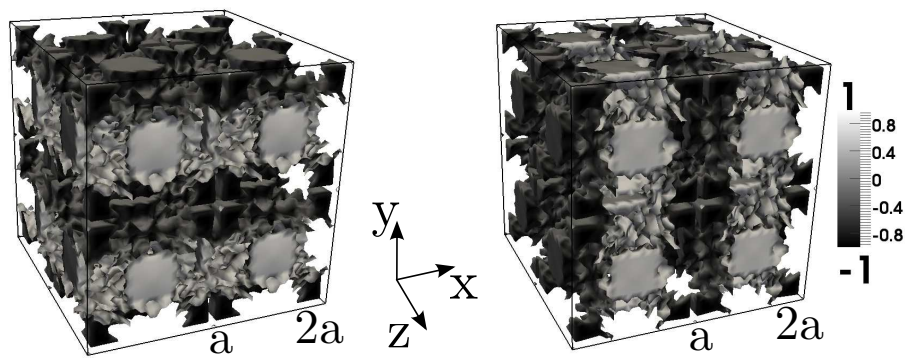}
\caption{Energy landscape at $T=1$ in a cubic box of $2a\times 2a\times 2a$
dimension where $a$ is the lattice parameter,
for the first and second degenerate configurations (left and right, respectively). The energy scale is indicated on the figure.}
\label{lands}
\end{center}
\end{figure}

 Let us show now how the itinerant spins choose their paths to travel across the lattice.  We show in Fig. \ref{lands} the energy landscape at $T=1$ for both degenerate configurations.  This gives some information concerning the spatial energy distribution in the system.


\begin{figure}
\begin{center}
\includegraphics[width=9cm]{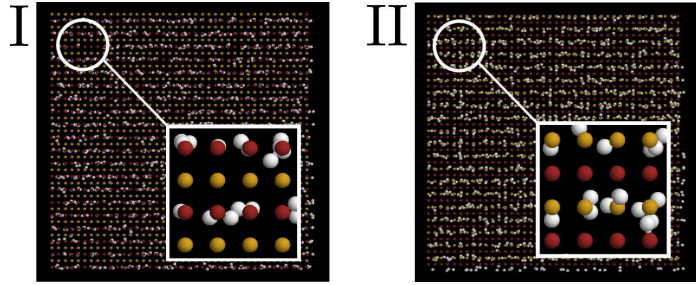}
\caption{(Color on line) 3D antiferromagnetic FCC lattice at $T=1$,
where lattice down spins are presented in red, up spins
in yellow and itinerant spins (which are up) in white.
The plane shown is the $xy$ plane with
$x$ direction (spin flow direction) being horizontal. Left: snapshot in the case $D_1=a$. Right: snapshot in the case $D_1=1.4a$.}
\label{3Dstruct}
\end{center}
\end{figure}

As we said above, the spin motion depends solely on the spin energy due to its interaction with surrounding spins.  The lower energy it has the longer it stays in that position.
By examining the different traveling paths we come to this observation: for a given $D_1$, the itinerant spin will choose its path where its energy is low. This is understandable from a viewpoint of statistical physics.   Paradoxically, by choosing low-energy paths, its motion is slowed down because as said earlier itinerant spins feel energetically at ease so it does not want to move.  So, depending on the value of  $D_1$,  itinerant spins will move near up-spin planes or near down-spin planes in order to have a low energy (see Fig. \ref{3Dstruct}).

\begin{figure}[!h]
\begin{center}
\includegraphics[width=12cm]{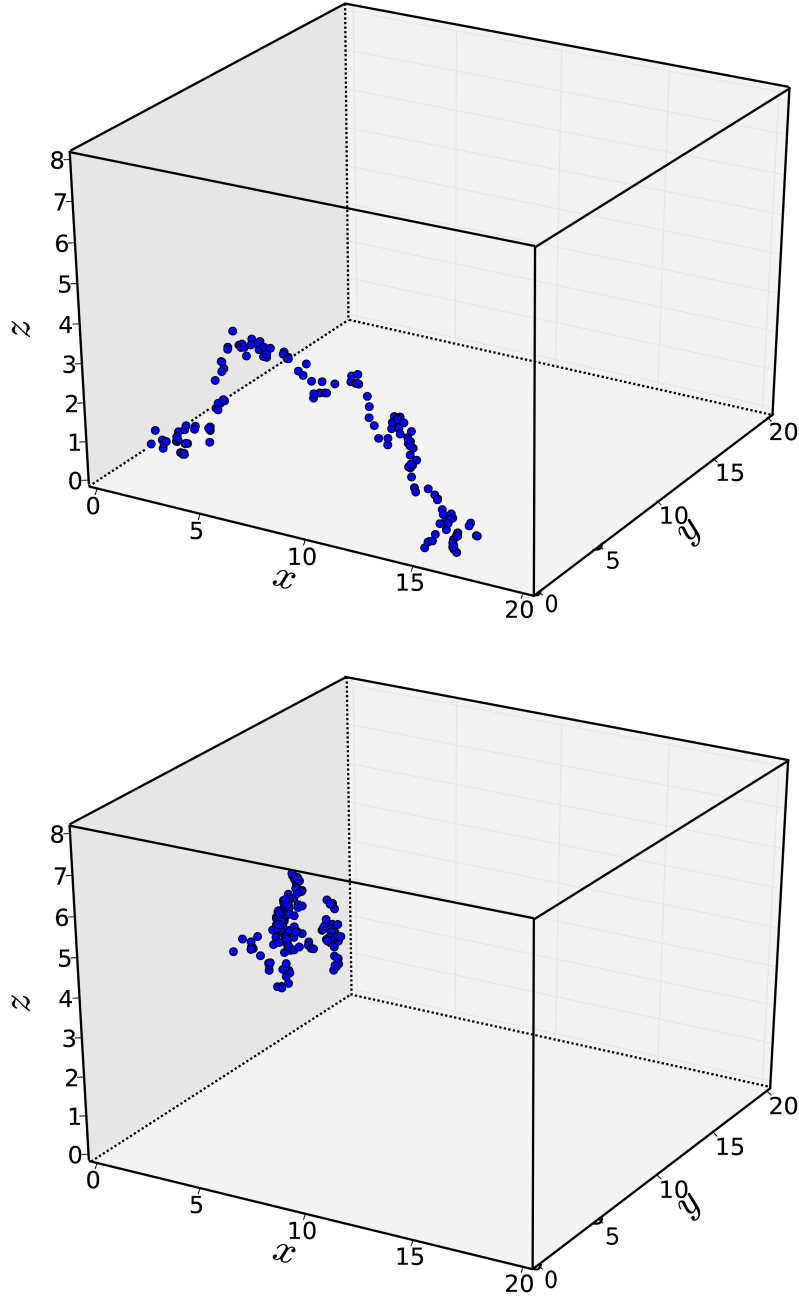}
\caption{Travel path of an itinerant spin at $T=1$ in the
first degenerate state (upper) and in the second degenerate
state (lower) during an equal travel time with $D_1=a$.  As seen, spins move more easily in the first degenerate configuration than in the second one. Other parameters are  $N_0=1600$,
$I_0=K_0=0.5$, $D=0.35$, $J_s=J=-1.0$}
\label{path}
\end{center}
\end{figure}

Let us show now in Fig. \ref{path} how the spins travel across the system. As seen, for equal travel time an itinerant spin moves faster and farther in the first degenerate configuration than in the second one for $D_1=a$.  This can be understood because the itinerant (up) spin is stopped for a more or less long time in front of a wall of down spins perpendicular to its $x$ trajectory in the second degenerate configuration.  However, when $D_1$ is very large, for instance 1.4$a$, there is no more difference between the two configurations because the distance is long enough for an itinerant spin to see other spins across down-spin walls.

\subsection{Discussion}

Let us recall that in ferromagnets, $\rho$ shows a peak at $T_C$. This peak was interpreted as a consequence of spin-spin
correlation.  The form of the peak depends on the correlation range\cite{DeGennes,Fisher,Kataoka}.  Other interpretations which were based
on scattering by defect clusters\cite{Akabli2} or by impurities\cite{Zarand} are all in agreement.   In antiferromagnets, the situation is quite different.
Unlike in ferromagnets where itinerant spins are slowed down only when they encounter antiparallel spins of defect clusters, in antiferromagnets itinerant spins see both parallel and antiparallel spins in any of its position, so their motion  depends drastically on their immediate local spin configuration whose energy is determined by the interaction range $D_1$.  In addition,
the behavior of the spin resistivity  in antiferromagnets depends on several other ingredients among which one can mention the crystal structure, the nature of the magnetic ordering, and the spin model.  We have simulated some non frustrated antiferromagnets such as Ising antiferromagnetic simple cubic (SC) and body-centered cubic (BCC) lattices. The spin resistivity shows no peak in these cases\cite{Magnin}.



\section{Results for the Heisenberg case}\label{Heis}

In this section, we presently briefly the results on the same lattice with the Heisenberg spin model. Itinerant spins are the same as used above, namely polarized Ising spins. This assumption allows to outline only the effect of the continuous nature of the Heisenberg lattice spin on the resistivity.  The full Hamiltonian with different kinds of interaction  is assumed as above except the exchange interaction between lattice spins. This is given by
\begin{equation}
\mathcal H= -\sum_{\left<i,j\right>}J_{i,j}\mathbf S_i\cdot\mathbf S_j -A\sum_{\left<i,j\right>} S^z_i S^z_j
\label{HamilR-H}
\end{equation}
where $\mathbf S_i$ is the Heisenberg spin at the site $i$ and $A$ an Ising-like anisotropy which is assumed to be negative to favor an antiparallel spin ordering on the $z$ axis.
When $A$ is zero, one has the isotropic Heisenberg model. In order to have at phase transition at a nonzero $T$, we should take a nonzero value for $A$ because it is known, by the theorem of Mermin-Wagner\cite{Mermin}, that for vector spin models there is no long-ranged ordering at finite temperatures in two dimensions.  The small thickness considered here is, in a phase-transition point of view, equivalent to a two dimensional system.  Except $A$, note that we use the same assumptions as in Eq. (\ref{HamilR}).

\begin{figure}[!h]
\begin{center}
\includegraphics[width=12cm]{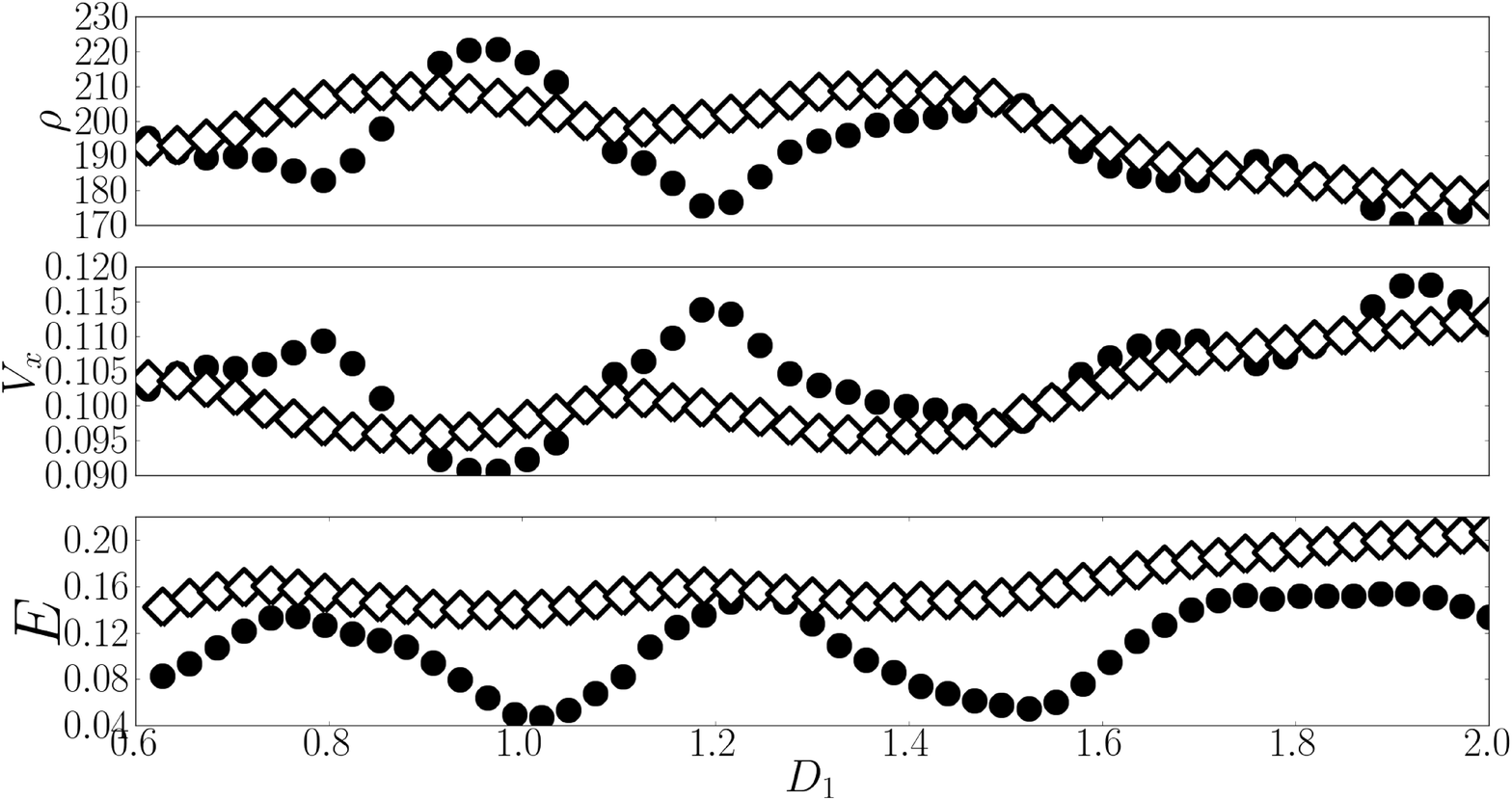}
\includegraphics[width=12cm]{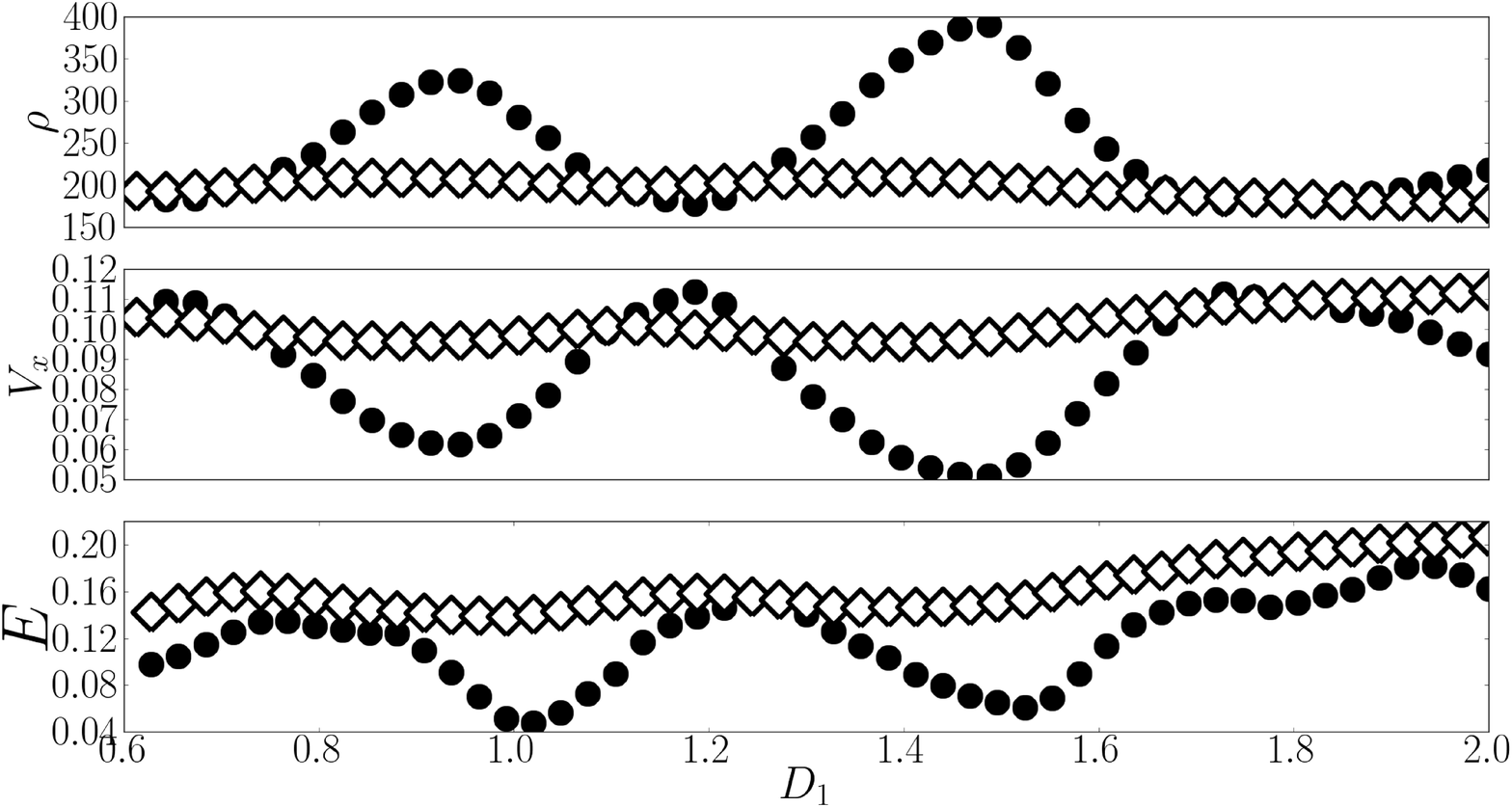}
\caption{Heisenberg case. Resistivity, spin velocity, and energy of itinerant spin versus $D_1$ (in unit of the lattice constant $a$)at temperatures
 $T=0.75$ (black circles) and $T=0.85$ (white diamonds) for first (upper) and second (lower) degenerate
configurations. $A=-1$, $N_z=8$, $N_0=1600$, $J_s=J=-1.0$.}
\label{Heis-RD1-DG2}
\end{center}
\end{figure}

The transition temperature with $A=-1$ is $T_C \simeq 0.79$ for the lattice size $N_x=N_y=20$, $N_z=8$.  We use here the same analysis as for the Ising case above: we first look at the effect of $D_1$ on the resistivity at two temperatures,  one lower and one higher than $T_C$.  This is shown in Fig. \ref{Heis-RD1-DG2} where the upper (lower) figure is for the first (second) degenerate spin configuration.
Again here, one observes that the two degenerate states do not yield the same transport properties as in the Ising case. The same remarks on physical mechanism are thus applied (see \ref{IDG1} and \ref{IDG1}).


\begin{figure}[!h]
\begin{center}
\includegraphics[width=12cm]{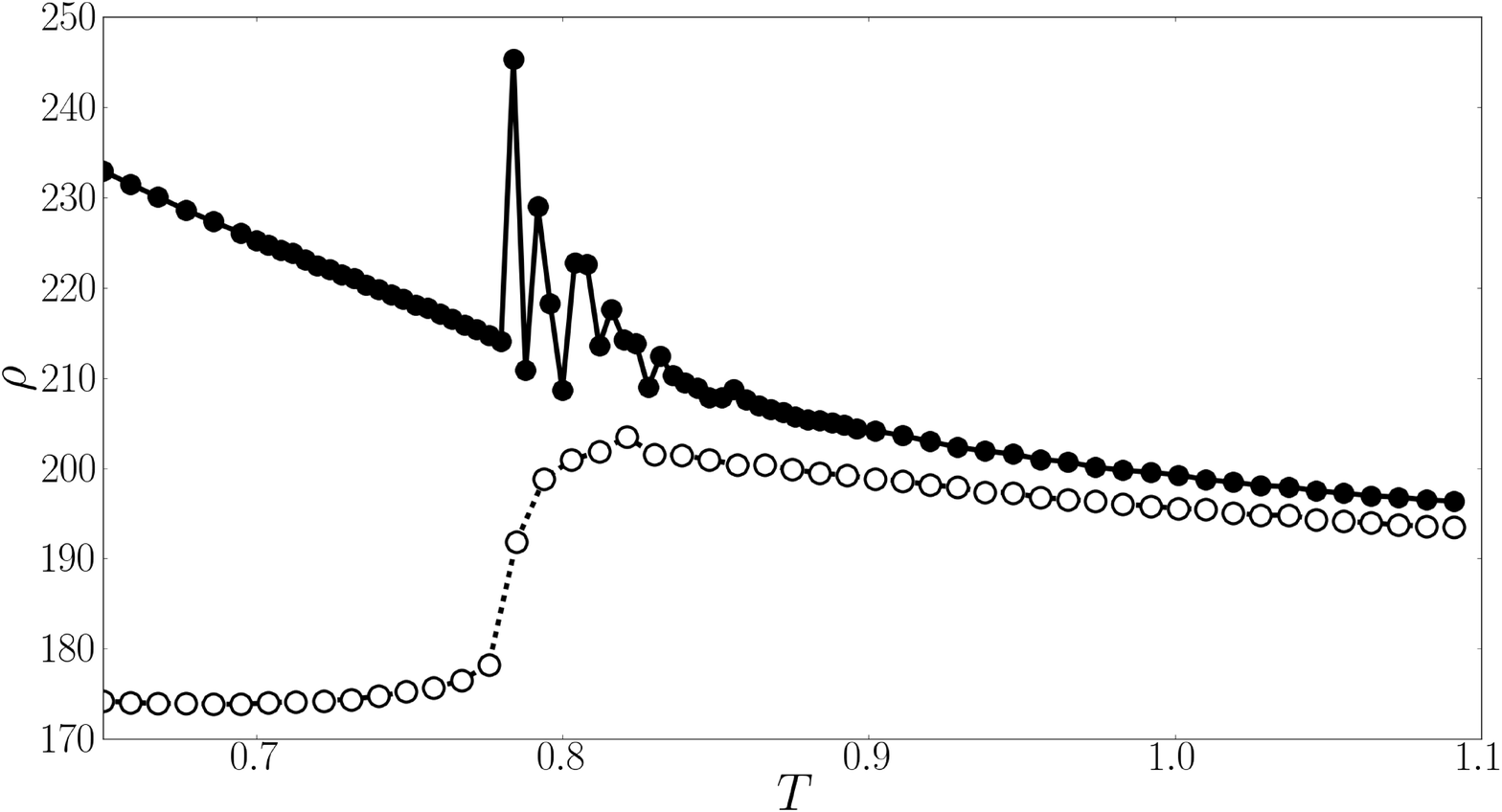}
\includegraphics[width=12cm]{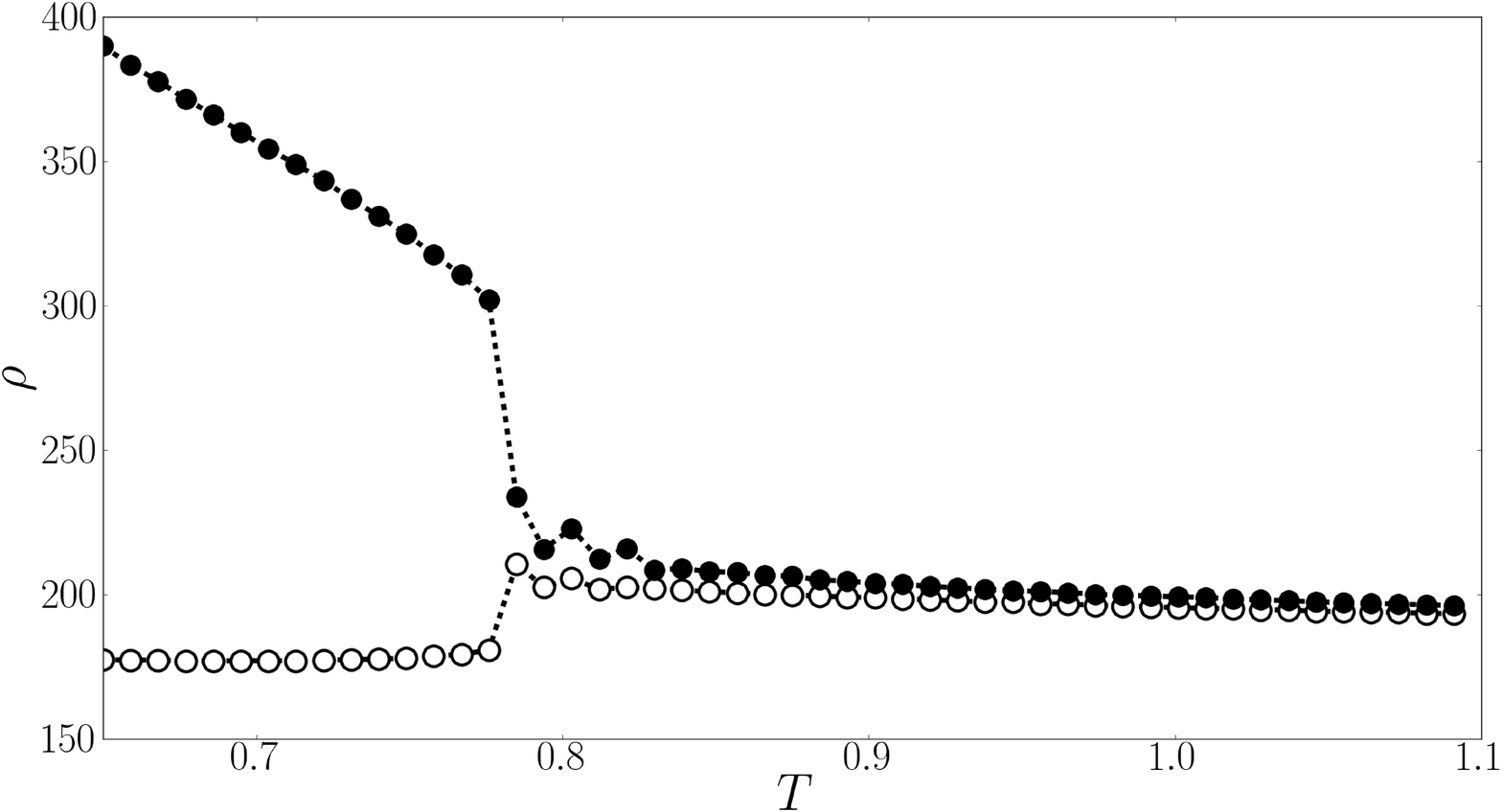}
\caption{Heisenberg case. Resistivity of thin film of size  $N_x=N_y=20$ and $N_z= 8$ for $N_0=1600$
itinerant spins versus $T$ for $D_1=a$ (black circles) and $D_1=1.25a$ (white circles) in unit of the lattice constant $a$ for first (upper) and second (lower) degenerate
states. $A=-1$, $J_s=J=-1.0$, $I_0=K_0=0.5$, $D=0.35$.}
\label{Heis-r_plot}
\end{center}
\end{figure}

Let us show now in Fig. \ref{Heis-r_plot} the resistivity as a function of $T$ for two typical values of $D_1$.   As seen, depending on the value of $D_1$, $\rho$ undergoes a sharp increase or decrease at $T_C$. At some values such as that corresponding to the upper curve of the upper figure, the resistivity can go across a large region of fluctuations without a sharp jump.  So in experiments, care should be taken to interpret similar behavior if any.
Note that the second degenerate configuration yields always a larger resistivity than in the first one, as observed in the Ising case in the previous section.

The effect of $A$ on the resistivity is not very important in the reasonable range $[0.1,1.5]$: except the fact that $T_C$ varies with $A$, for instance $T_C\simeq 0.65$ for $A=0.5$ and $T_C\simeq 0.55$ for $A=0.1$, the discontinuity of $\rho$ at $T_C$ diminishes only slightly with decreasing $A$.


\section{Conclusion}
The model used in this paper is rather general. It has been applied  to ferromagnets with success\cite{Akabli2}.  We believe that it can be applied to different materials by choosing appropriate interactions. For example, in metals we have to reduce to almost zero the interaction between itinerant spins and lattice spins
to create the situation of almost-free conduction electrons.  The lattice disordering transition in magnetic metals then should not strongly affect the spin resistivity. In semiconductors, that interaction should be strong enough to reproduce a peak of $\rho$ as experimentally observed.

We have shown in this paper that the spin resistivity  $\rho$ of the fully frustrated FCC antiferromagnet is quite different from that of ferromagnets\cite{Akabli2} and non frustrated antiferromagnets\cite{Magnin}.
$\rho$  does not show a peak at the magnetic phase transition temperature.  It shows instead a discontinuous jump at the transition temperature $T_C$. The jump depends on the numbers of parallel and antiparallel localized spins which interact with an itinerant spin.   After transition, the resistivity tends to  a saturation value independent of $D_1$.
The abrupt behavior of $\rho$ at $T_C$ in the AF FCC Ising lattice is an effect of the frustration which causes a first-order transition of the lattice magnetic ordering leading to a discontinuity of $\rho$ at $T_C$.

We are not aware of experiments performed on spin transport in  materials with first-order magnetic transition. Our result is thus a prediction which would be useful for future experiments.  Note however that for electrical transport, the electrical resistivity shows a discontinuity at a metal-insulator "first-order" transition in PrNi$O_3$\cite{Granados} and NdNi$O_3$\cite{Granados93}.  Our magnetic resistivity found in this paper has also a discontinuity behavior at a magnetic "first-order" transition.  This similarity shows that the resistivity is closely related to the nature of the phase transition, whatever its origin (magnetic, insulator-metal, ...) may be.   The mapping between the two cases however is not the scope of this paper.

We have also shown that the surface disordering causes a peak of the resistivity at the surface transition temperature.    In the Heisenberg model, the spin continuous degrees of freedom weaken the first-order transition, yielding in general a reduction of the critical temperature and a less abrupt change of the resistivity at the transition.

As a last remark, let us emphasize that the behavior of the spin resistivity at $T_C$ is quite different from one antiferromagnet to another. It depends on many factors such as the lattice structure, the interaction range, the spin model and the instability (in particular due to frustration) of the spin ordering. We have studied here the effects of some of them, but a throughout understanding needs much more investigations and analysis.

KA acknowledges a financial support from the JSPS for his stay at Okayama University.  He wishes to thank Prof. I. Harada for helpful discussion.

{}
\end{document}